\documentclass[prx,preprint,superscriptaddress]{revtex4-1}
\usepackage{amsfonts}
\usepackage{amsmath}
\usepackage{amssymb}
\usepackage{lineno}
\usepackage{color}
\usepackage{graphicx}%
\usepackage{multirow}
\begin{document}


\title{Anomalous thermal properties and spin crossover of ferromagnesite (Mg,Fe)CO$_3$}

\author{Han Hsu}
\email[Corresponding author: ]{hanhsu@ncu.edu.tw}
\affiliation{Department of Physics, National Central University, Taoyuan City 32001, Taiwan}
\author{C. Crisostomo}
\affiliation{Department of Physics, National Central University, Taoyuan City 32001, Taiwan}
\author{Wenzhong Wang}
\affiliation{School of Earth and Space Sciences, University of Science and Technology of China, Hefei, Anhui 230026, P.~R. China}
\author{Zhongqing Wu}
\affiliation{School of Earth and Space Sciences, University of Science and Technology of China, Hefei, Anhui 230026, P.~R. China}

\date{\today}

\begin{abstract}

Ferromagnesite [(Mg$_{1-x}$Fe$_{x}$)CO$_3$], also referred to as magnesiosiderite at high iron concentration ($x>0.5$), is a solid solution of magnesite (MgCO$_3$) and siderite (FeCO$_3$). Ferromagnesite is believed to enter the Earth's lower mantle via subduction and is considered a major carbon carrier in the Earth’s lower mantle, playing a key role in the Earth's deep carbon cycle. Experiments have shown that ferromagnesite undergoes a pressure-induced spin crossover, accompanied by volume and elastic anomalies, in the lower-mantle pressure range. In this work, we investigate thermal properties of (Mg$_{1-x}$Fe$_x$)CO$_3$ ($0<x \leq 1$) using first-principles calculations. We show that nearly all thermal properties of ferromagnesite are drastically altered by iron spin crossover, including anomalous reduction of volume, anomalous softening of bulk modulus, and anomalous increases of thermal expansion, heat capacity, and Gr\"uneisen parameter. Remarkably, the anomaly of heat capacity remains prominent (up to $\sim 40$\%) at high temperature without smearing out, which suggests that iron spin crossover may significantly affect the thermal properties of subducting slabs and the Earth's deep carbon cycle.
 
\end{abstract}

\maketitle

\section{Introduction}

Ferromagnesite [(Mg$_{1-x}$Fe$_{x}$)CO$_3$], also referred to as magnesiosiderite at high iron concentration ($x>0.5$), is a solid solution of magnesite (MgCO$_3$) and siderite (FeCO$_3$), both crystallizing in $R\bar3c$ symmetry (space group No.~167) at ambient conditions. Ferromagnesite is believed to enter the Earth's lower mantle (660--2890 km deep, pressure range 23--135 GPa) via subduction and is considered a major carbon carrier in the Earth’s lower mantle, playing a key role in the Earth's deep carbon cycle \cite{Hazen13, Manning20}. Experiments have shown that ferromagnesite remains stable up to 115 GPa and 1300--3000 K (depending on pressure, iron concentration, and iron spin state) \cite{Isshiki04, Liu_2015_SciRep, Cerantola_2017_NatCommun}. Beyond the above-mentioned pressure ($P$) and temperature ($T$) range, ferromagnesite undergoes various complicated structural transitions and redox reactions, depending on iron concentration. Orthorhombic, monoclinic, and triclinic phases of (Mg,Fe)CO$_3$, and Fe$^{3+}$-bering Mg$_2$Fe$_2$C$_4$O$_{13}$, Fe$_4$C$_4$O$_{13}$, and Fe$_4$C$_3$O$_{12}$ have been proposed based on experiments \cite{Isshiki04, Liu_2015_SciRep, Cerantola_2017_NatCommun, Boulard11, Boulard15, Merlini15, Boulard20} and first-principles calculations \cite{Oganov08, Pickard15, Li20, Tsuchiya20}, and consensus has not been reached. Clearly, iron directly affects the properties of (Fe,Mg)-bearing carbonates, including their structural transitions and phase boundaries.  

\bigskip

One more complexity of ferromagnesite arises from iron spin crossover (SCO), also referred to as spin transition: The total electron spin ($S^{el}$) of iron varies with pressure and temperature. At ambient conditions, Fe$^{2+}$ in ferromagnesite adopts the high-spin (HS, $S^{el}=2$) state; upon compression, $S^{el}$ decreases. Signatures of SCO in ferromagnesite have been observed via various spectroscopic techniques, including x-ray emission \cite{Mattila07} and absorption \cite{Cerantola15}, M\"ossbauer \cite{Cerantola15}, Raman \cite{Farfan12, Lin12, Cerantola15, Muller16, Muller17, Weis17, Chao_2019_JGR}, and optical absorption spectroscopy \cite{Lobanov15, Lobanov16, Taran19}. In addition, volume and elastic anomalies accompanying SCO have been observed via x-ray diffraction \cite{Farfan12, Lin12, Lavina09, Lavina10HPR, Lavina_2010_PRB, Nagai10, Liu_2014_AmMin} and Brillouin scattering \cite{Stekiel17, Fu_2017_PRL}, respectively. In the above-mentioned room-temperature ($T=300$ K) experiments, SCO typically starts at 40--49 GPa and finishes at 46--56 GPa; the typical width of the SCO region is 5--10 GPa. An exception is observed via M\"ossbauer spectroscopy, indicating an SCO region of 52--61 GPa \cite{Cerantola15}. Our previous \textit{static} calculation has confirmed that only the HS and the low-spin (LS, $S^{el}=0$) states are involved in the SCO of ferromagnesite, while the intermediate-spin (IS, $S^{el}=1$) state is highly unlikely \cite{Hsu_2016_PRB}. The HS--LS SCO region and volume anomaly given by our calculation are also in good agreement with experiments \cite{Hsu_2016_PRB}. So far, most experimental studies for the SCO of ferromagnesite are conduced at room temperature. Studies for the thermal properties of ferromagnesite at high $P$--$T$ conditions have been scarce \cite{Liu_2014_AmMin}, despite the necessity of high $P$--$T$ experiments to fully understand the SCO of ferromagnesite in the Earth's interior and its potential geophysical and geochemical effects.

\bigskip

In a broader perspective, iron is incorporated in many minerals in the Earth's interior, including ferropericlase [(Mg,Fe)O] and Fe-bearing bridgmanite (MgSiO$_3$ perovskite), which constitute $\sim 20$ and $\sim 75$ vol\% of the Earth's lower mantle, respectively. Extensive studies on these two minerals have shown that SCO directly affects the physical properties of the host minerals and also affects iron diffusion and partitioning in the Earth's interior (see Refs.~\onlinecite{Hsu_2010_Rev, Wentzcovitch_2012_Rev, Lin_2013_Rev, Badro_2014_Rev} for review). SCO of ferropericlase is now propsed to control the structure of the large low velocity provinces \cite{Huang15} and to generate the anticorrelation between bulk sound and shear velocities in the lower mantle \cite{Wu14}. Further geophysical and geochemical effects of SCO have been anticipated \cite{Lin_2013_Rev, Badro_2014_Rev}. In addition to ferropericlase, bridgmanite, and ferromagnesite, a few more minerals of potential geophysical and geochemical importance have also been reported to undergo SCO, including Fe-bearing new hexagonal aluminous (NAL) phase NaMg$_2$(Si,Al)$_6$O$_{12}$ \cite{Wu_2016_EPSL, Lobanov_2017_JGR, Hsu_2017_PRB}, calcium-ferrite aluminous (CF) phase (Na,Mg)(Si,Al)$_2$O$_4$ \cite{Wu17}, and pyrite-type FeO$_2$H$_y$ ($0 \leq y \leq 1$)  \cite{Liu19, Jang19}. While plenty of mantle minerals are subject to SCO, studies for their thermal properties during SCO at high $P$--$T$ conditions have been scarce. Recently, anomalous changes of thermal conductivity during SCO have been observed in ferromagnesite \cite{Chao_2019_JGR}, ferropericlase \cite{Ohta17, Hsieh18}, and bridgmanite \cite{Hsieh17, Okuda19} via pulsed light heating thermoreflectance and time-domain thermoreflectance (TDTR) experiments. In these TR-based experiments, either thermal diffusivity \cite{Ohta17, Okuda19} or thermal effusivity \cite{Hsieh18, Hsieh17} are measured. To extract thermal conductivity from TR-based experiments, heat capacity is a necessary input \cite{Hasegawa19, Jiang18}. In practice, since heat capacities at high $P$--$T$ conditions are not easily available, estimated values are often adopted \cite{Hsieh18, Hsieh17}. This approach, however, may lead to inaccurate estimate of thermal conductivity, as the anomalous change of heat capacity during SCO (see Sec. III and Ref.~\onlinecite{Wu_2009_PRB}) is ignored. A comprehensive computational study is thus desirable, to provide necessary information for the analysis of TR-based experiments, and to further shed light on the thermal properties of ferromagnesite and related materials during SCO at high $P$--$T$ conditions.

\section{Computational Method}

\begin{figure}[pt]
\begin{center}
\includegraphics[
]{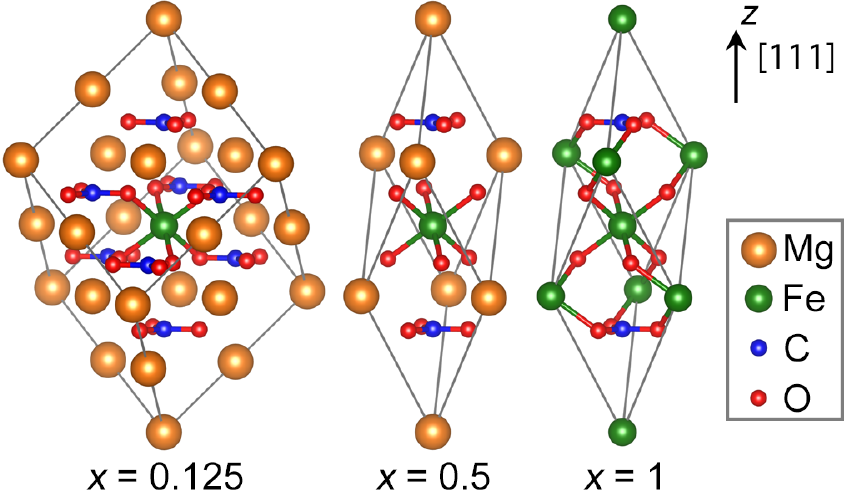}
\end{center}
\caption{Atomic structures of ferromagnesite (Mg$_{1-x}$Fe$_x$)CO$_3$ for $x=0.125$ (40-atom supercell) and $x=0.5$ and $1$ (10-atom cell). The end member FeCO$_3$ ($x = 1$) crystalizes in calcite structure ($R\bar3c$ symmetry), same as MgCO$_3$ ($x=0$, not shown). In this graph, the [111] direction is aligned with the $z$ axis.} 
\label{Fig:unitcells}
\end{figure}

In this work, all calculations are performed using the {\sc Quantum ESPRESSO} codes \cite{PWscf2017}; ultrasoft pseudopotentials (USPPs) generated with the Vanderbilt method \cite{Vanderbilt90} are adopted. To properly treat the on-site Coulomb interaction of Fe-$3d$ electrons, we use the local density approximation + self-consistent Hubbard $U$ (LDA+$U_{sc}$) method, with the $U$ parameters computed self-consistently \cite{Cococcioni_2005_PRB, Kulik_2006_PRL, Himmetoglu_2011_PRB, Himmetoglu_2014_IJQC}. Via LDA+$U_{sc}$ calculations, SCO (or the lack thereof) in ferropericlase, bridgmanite, MgSiO$_3$ post-perovskite, ferromagnesite, and the NAL phase have been successfully elucidated \cite{Hsu_2016_PRB, Hsu_2017_PRB, Hsu_2010_EPSL, Hsu_2011_PRL, Yu_2012_EPSL, Hsu_2012_EPSL, Hsu_2014_PRB}. Here we adopt the previously reported $U_{sc}=4.0$ and $5.4$ eV for the HS and LS Fe$^{2+}$, respectively \cite{Hsu_2016_PRB}. Structural optimizations for (Mg$_{1-x}$Fe$_x$)CO$_3$ with $x=0.125$ and with $x=0.5$ or $1$ are performed using 40 and 10-atom cells, respectively, as shown in Fig.~\ref{Fig:unitcells}. Phonon calculations are performed using the Phonopy package, in which finite-displacement method is implemented \cite{Phonopy}. Within this method, we adopt supercells containing up to 270 (for $x=0.5$ and $1$) or 320 (for $x=0.125$) atoms. With the phonon spectra $\omega_{\nu\bold q}^{i}(V)$ of spin state $i$ ($i=$ HS, IS, or LS) at volume $V$ obtained, we compute the vibrational free energy $F^{vib}_i(T,V)$ within the quasi-harmonic approximation (QHA); the equation of state $V_i(P,T)$, Gibbs free energy $G_i(P,T)$, and other thermal parameters of spin state $i$ can be determined accordingly, as detailed in Supplemental Material (SM) \cite{SM}. We fit our calculation results with the third-order Birch--Murnaghan equation of state (3rd BM EoS) using the qha Python package \cite{qha}.

\bigskip

At nonzero temperatures ($T \neq 0$), ferromagnesite goes through a mixed-spin (MS) phase/state, in which all spin states coexist. The fraction of spin state $i$ in the MS phase is written as $n_i=n_i(P,T)$. For ferromagnesite, the IS state is energetically \textit{unfavorable}, and the IS fraction $n_{IS}$ is negligible \cite{Hsu_2016_PRB}. Effectively, $n_{IS}=0$, and $n_{LS}+n_{HS}=1$. For convenience, we write $n_{LS} \equiv n$ and $n_{HS}=1-n$. Based on the thermodynamic model detailed in SM \cite{SM} (see also Refs.~\onlinecite{Hsu_2010_Rev, Wu_2009_PRB, Tsuchiya06}), the LS fraction $n(P,T)$ is given by

\begin{equation}
n = \frac{1}{1+\exp(\Delta G_{LS}/k_{B}Tx)},
\label{Eq:n}
\end{equation}

\bigskip
\noindent
where $\Delta G_{LS} \equiv G_{LS}-G_{HS}$. With known LS fraction, the Gibbs free energy $G(P,T)$ of the MS phase can be written, from which all thermal parameters of the MS phase can be derived (see SM \cite{SM}).

\section{Results}

\begin{figure}[pt]
\begin{center}
\includegraphics[
]{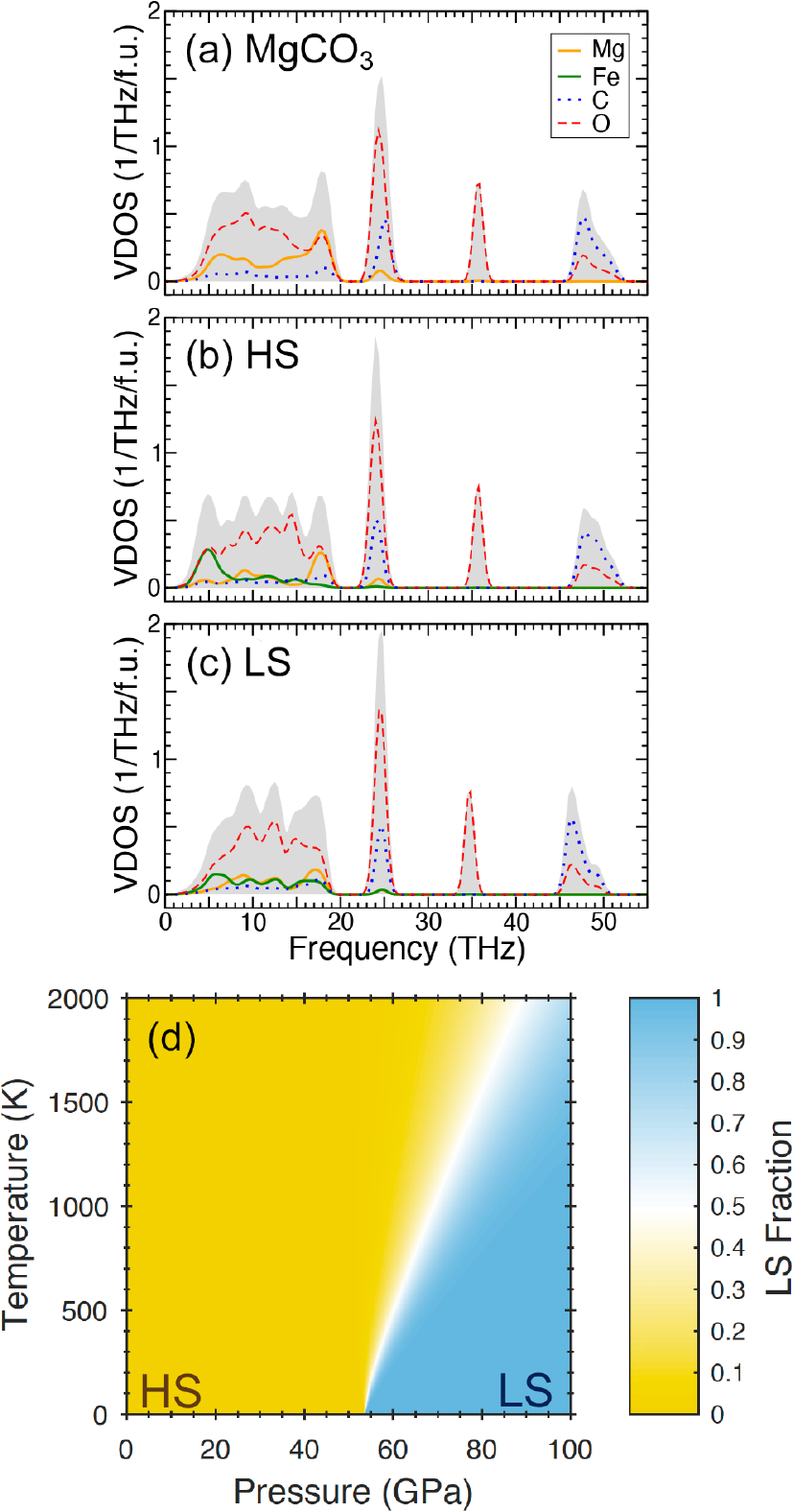}
\end{center}
\caption{Vibrational density of states of ferromagnesite (Mg$_{1-x}$Fe$_x$)CO$_3$ at volume $V=37.01$ \AA$^3$/f.u. for (a) $x=0$ and (b,c) $x=0.5$ in the HS and LS states, respectively. In panels (a)--(c), the gray shades denote the total VDOS; the lines denote the projected VDOS onto the Mg, Fe, C, and O atoms. Also, $1$ THz $=33.356$ cm$^{-1}$. (d) Spin phase diagram of (Mg$_{0.5}$Fe$_{0.5}$)CO$_3$; the fraction of LS iron is indicated by color.}
\label{Fig:VDOS_nPT}
\end{figure}

\begin{figure}[pt]
\begin{center}
\includegraphics[
]{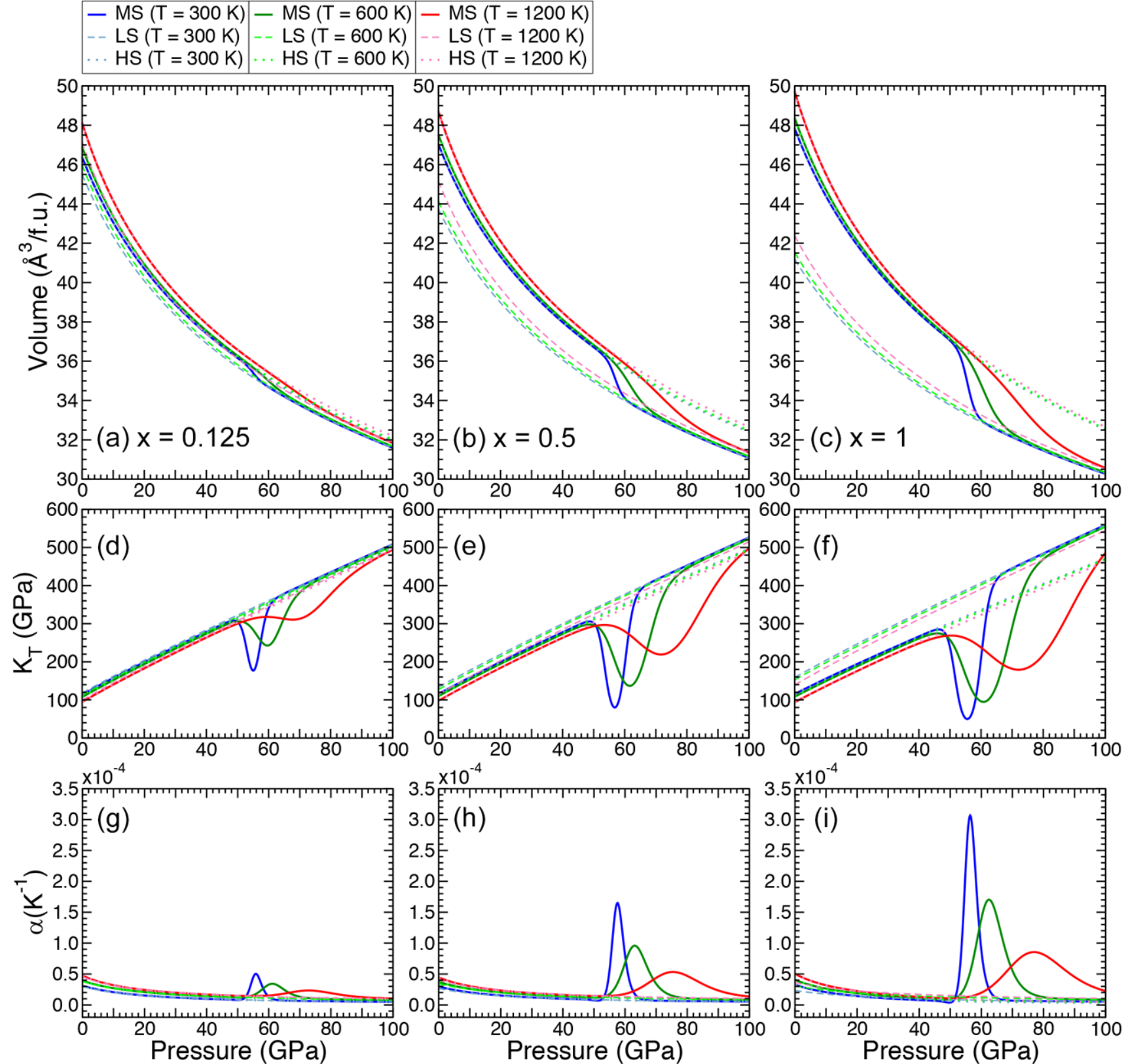}
\end{center}
\caption{(a--c) Compression curves $V(P)$, (d--f) isothermal bulk modulus $K_T$, and (g--i) volumetric thermal expansivity $\alpha$ of (Mg$_{1-x}$Fe$_x$)CO$_3$ for $x=0.125$, $0.5$, and $1$, respectively. Solid, dotted, and dashed lines denote our theoretical results for the MS, HS, and LS states, respectively.}
\label{Fig:V_Kt_alpha}
\end{figure}

\begin{figure}[pt]
\begin{center}
\includegraphics[
]{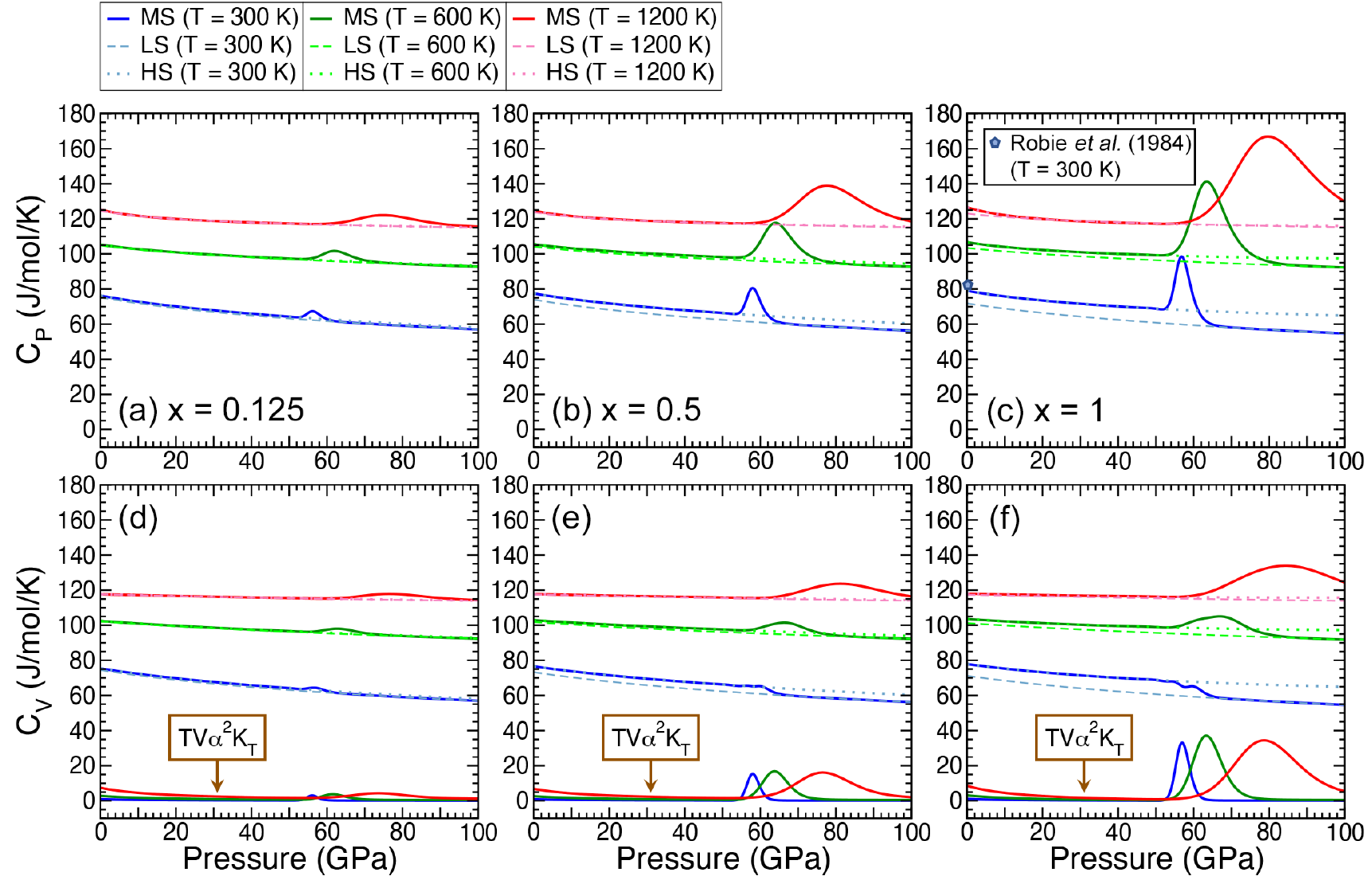}
\end{center}
\caption{(a--c) Constant-pressure ($C_P$) and (d--f) constant-volume ($C_V$) heat capacities of (Mg$_{1-x}$Fe$_x$)CO$_3$ for $x=0.125$, $0.5$, and $1$, respectively. The differences between $C_P$ and $C_V$, i.e. $TV\alpha^2K_T$, are also shown in panels (d)--(f). Solid, dotted, and dashed lines denote our theoretical results for the MS, HS, and LS states, respectively; symbols in panel (c) denote experimental results \cite{Robie_1984_AmMin}.}
\label{Fig:Cp_Cv}
\end{figure}

To analyze the lattice vibration of (Mg$_{1-x}$Fe$_x$)CO$_3$, we plot the vibrational density of states (VDOS) of MgCO$_3$ and (Mg$_{0.5}$Fe$_{0.5}$)CO$_3$ at $V=37.01$ \AA$^3$/f.u., as shown in Fig.~\ref{Fig:VDOS_nPT}. At this volume, the vibrational frequencies of Mg atoms are in the region of 0--20 THz (0--667 cm$^{-1}$) [Fig.~\ref{Fig:VDOS_nPT}(a)]; HS Fe atoms vibrate with frequencies of 3--8 THz (100--267 cm$^{-1}$) [Fig.~\ref{Fig:VDOS_nPT}(b)]; LS Fe atoms vibrate with frequencies of 0--20 THz (0--667 cm$^{-1}$) [Fig.~\ref{Fig:VDOS_nPT}(c)]. The lower average vibrational frequency of HS Fe compared to LS Fe arises from the smaller interatomic force constants (IFCs) between HS Fe and neighboring atoms. At $V=37.01$ \AA$^3$/f.u., the mean force constants of HS and LS Fe are $328.8$ and $549.9$ N/m, respectively. These results are consistent with the smaller bulk modulus and larger heat capacity of HS ferromagnesite (see discussions of Figs.~\ref{Fig:V_Kt_alpha} and \ref{Fig:Cp_Cv}). For ferromagnesite, only the HS and LS states can be observed \cite{Hsu_2016_PRB}; a spin phase diagram can be obtained by plotting the LS fraction $n(P,T)$. Figure \ref{Fig:VDOS_nPT}(d) is the spin phase diagram of (Mg$_{0.5}$Fe$_{0.5}$)CO$_3$, where the LS fraction is indicated by color. Here, we use the white color to indicate $n=0.5$, which is equivalent to $\Delta G_{LS} = 0$ [see Eq.~(\ref{Eq:n})]. The white color thus also marks the spin-transition pressure $P_t$ and the boundary between the HS and LS states. Evidently, (Mg$_{0.5}$Fe$_{0.5}$)CO$_3$ undergoes a sharp HS--LS transition with a very narrow SCO region at low temperature. As the temperature increases, the width of the SCO region is broadened, the sharp spin transition becomes a smoother and broader SCO, and the spin-transition pressure $P_t$ increases.

\bigskip

\begin{figure}[pt]
\begin{center}
\includegraphics[
]{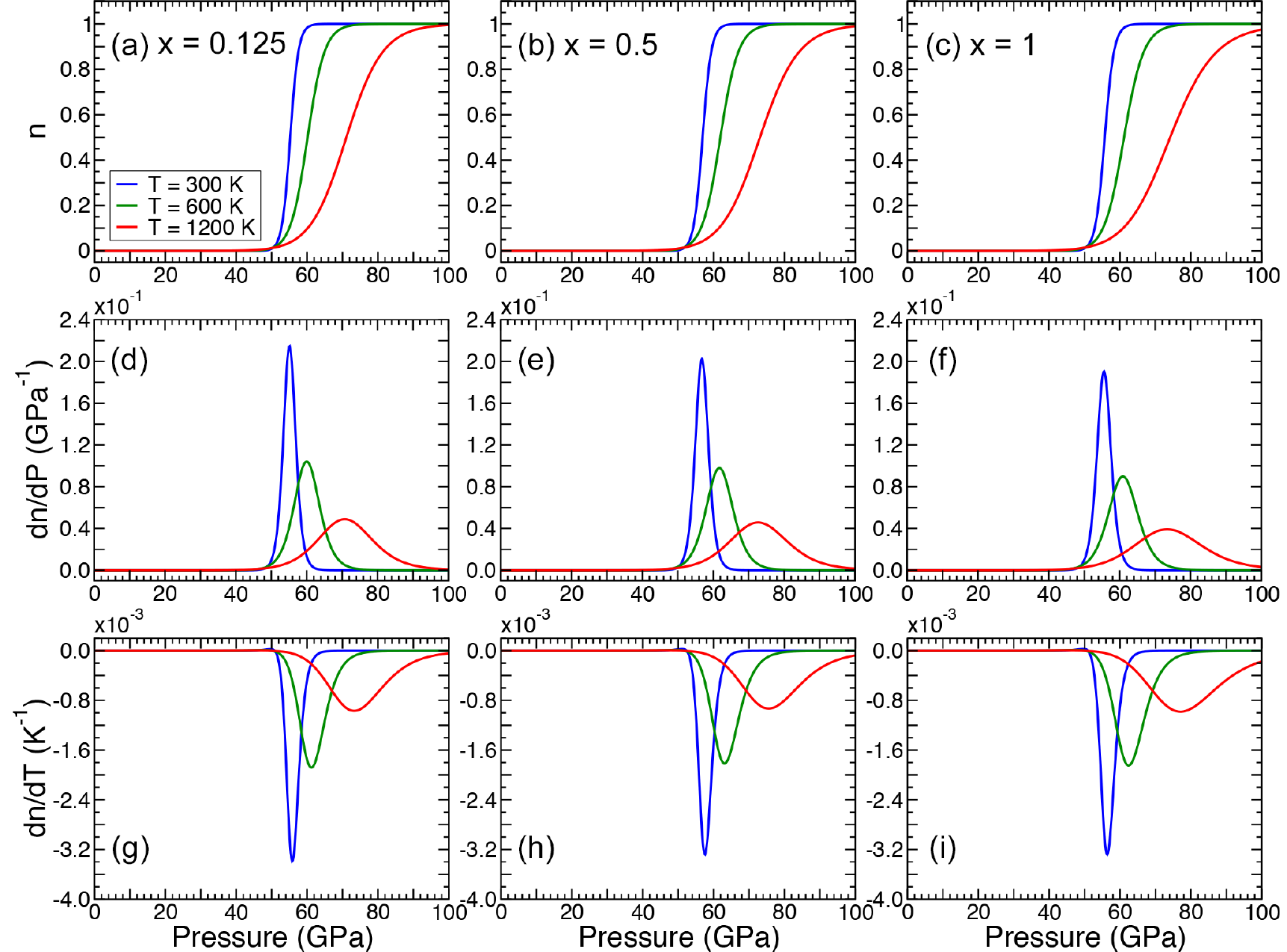}
\end{center}
\caption{(a--c) Fractions of LS iron ($n \equiv n_{LS}$) in (Mg$_{1-x}$Fe$_x$)CO$_3$ at various temperatures for $x=0.125$, $0.5$, and $1$, respectively; (d--f) $\partial n /\partial P$ and (g--i) $\partial n /\partial T$ for the Fe concentrations and temperatures considered in panels (a)--(c).}
\label{Fig:n_dndP_dndT}
\end{figure}

To better analyze the spin phase diagram of (Mg$_{1-x}$Fe$_x$)CO$_3$, we plot the isothermal LS fraction $n(P)$ for $T=300$, $600$, and $1200$ K in Fig.~\ref{Fig:n_dndP_dndT}. This choice of temperature is based on experimental results: For $P \gtrsim 50$ GPa, (Mg$_{1-x}$Fe$_x$)CO$_3$ with $x \geq 0.65$ is no longer stable at $T \gtrsim 1300$ K \cite{Liu_2015_SciRep, Cerantola_2017_NatCommun}. Here, we also investigate the effects of iron concentration by considering $x=0.125$, $0.5$, and $1$, as shown in Figs.~\ref{Fig:n_dndP_dndT}(a)--\ref{Fig:n_dndP_dndT}(c), respectively. Noticeably, the $n(P)$ curves for all three $x$'s are nearly the same, indicating that iron concentration barely affects the spin phase diagram. In contrast, for ferropericlase (Mg$_{1-x}$Fe$_x$)O, the spin-transition pressure $P_t$ significantly increases with $x$ when $0.25 \leq x \leq 1$ \cite{Lin_2013_Rev, Badro_2014_Rev}. Such difference arises from the stronger Fe-Fe interactions in (Mg$_{1-x}$Fe$_x$)O with $x>0.25$. In (Mg$_{1-x}$Fe$_x$)O, FeO$_{6}$ octahedra are corner-sharing when $x=0.25$ and can be edge- or face-sharing when $x>0.25$. Consequently, when $x>0.25$, Fe-Fe interactions are more significant, which affects the spin crossover. In contrast, in (Mg$_{1-x}$Fe$_x$)CO$_3$, FeO$_{6}$ octahedra are only corner-sharing even when $x=1$ (FeCO$_{3}$). Therefore, in (Mg$_{1-x}$Fe$_x$)CO$_3$, Fe-Fe interactions are weak, Fe atoms are effectively isolated from each other, and Fe concentration barely affects the spin crossover. Given such characteristic of (Mg$_{1-x}$Fe$_x$)CO$_3$, its SCO can be exemplified by the case of $x=0.5$ [Fig.~\ref{Fig:n_dndP_dndT}(b)]: At $T=300$, $600$, and $1200$ K, $P_t=57$, $62$, and $73$ GPa, and the widths of the SCO regions are $\sim 10$, $\sim 24$, and $\sim 45$ GPa, respectively. (Calculation results up to $T=2000$ K for $x=0.5$ are shown in SM \cite{SM}). Clearly, the computed $n(P)$ for $T=300$ K is in good agreement with room-temperature experiments reviewed in Sec.~I. In Figs.~\ref{Fig:n_dndP_dndT}(d)--\ref{Fig:n_dndP_dndT}(f) and ~\ref{Fig:n_dndP_dndT}(g)--\ref{Fig:n_dndP_dndT}(i), we plot the derivatives of $n(P,T)$ with respect to pressure and temperature, respectively, for their direct relevance to the anomalous changes of the bulk modulus and thermal expansivity, respectively, as shall be discussed later. Noticeably, as the temperature increases, the peaks of $\partial n / \partial P$ and the dips of $\partial n / \partial T$ are broadened, and their magnitudes are reduced. 

\bigskip


In Figs.~\ref{Fig:V_Kt_alpha}(a)--\ref{Fig:V_Kt_alpha}(c), we plot the compression curves $V(P)$ of (Mg$_{1-x}$Fe$_x$)CO$_3$ in the MS phase for iron concentrations $x=0.125$, $0.5$, and $1$, respectively. Compression curves of the pure HS and LS states ($V_{HS}$ and $V_{LS}$) are also plotted for reference; their EoS parameters ($V_0$, $K_0$, and $K_0'$) are tabulated in SM. As the iron concentration $x$ increases, $V_{HS}$ shifts up while $V_{LS}$ shifts down. This is because the ionic radius of the HS/LS Fe$^{2+}$ is larger/smaller than that of Mg$^{2+}$. By comparing $V(P)$ of the MS phase with the LS fraction $n(P)$ shown in Fig.~\ref{Fig:n_dndP_dndT}(a)--\ref{Fig:n_dndP_dndT}(c), one can notice that (1) before and after the SCO, $V(P)$ merges with $V_{HS}$ and $V_{LS}$, respectively, (2) anomalous volume reduction occurs during the SCO, and (3) volume anomaly and the SCO region are broadened by temperature. All these characteristics arise from $V(P)$ being the weighted average of $V_{LS}$ and $V_{HS}$ (see also Eq.~(S17) in SM \cite{SM}):

\begin{equation}
V(P) = \left(\frac{\partial G}{\partial P}\right)_T = nV_{LS}+(1-n)V_{HS},
\label{Eq:V}
\end{equation}

\bigskip
\noindent
which clearly indicates that the volume anomaly is directly related to the LS fraction $n$.

\bigskip

In Figs.~\ref{Fig:V_Kt_alpha}(d)--\ref{Fig:V_Kt_alpha}(f), we plot the isothermal bulk modulus $K_T \equiv -V(\partial P / \partial V)_T$ of the MS phase, along with its HS and LS counterparts ($K_{T}^{HS}$ and $K_{T}^{LS}$). For all iron concentrations and all temperatures, $K_{T}^{HS} < K_{T}^{LS}$, due to the smaller IFCs between the HS Fe and neighboring atoms [see  Fig.~\ref{Fig:VDOS_nPT}(b) and \ref{Fig:VDOS_nPT}(c)]. During the SCO, $K_T$ goes through an anomalous softening rather than just shift from $K_T^{HS}$ to $K_T^{LS}$. This can be understood via Eq.~(\ref{Eq:Kt}) below (see also Eq.~(S19) in SM \cite{SM}):

\begin{equation}
\frac{V}{K_T} = n \frac{V_{LS}}{K_{T}^{LS}} + (1-n) \frac{V_{HS}}{K_{T}^{HS}} + (V_{HS}-V_{LS})\left(\frac{\partial n}{\partial P} \right)_T,
\label{Eq:Kt} 
\end{equation}

\bigskip
\noindent
which indicates that the anomaly of $K_T$ mainly arises from $(\partial n/\partial P)_T$. By comparing $K_T$ with $(\partial n/\partial P)_T$ shown in Figs.~\ref{Fig:n_dndP_dndT}(d)--\ref{Fig:n_dndP_dndT}(f), one can notice that the peaks of ($\partial n/\partial P)_T$ and the dips of $K_T$ not only align with each other, but are also broadened and smeared by temperature in the same manner. Likewise, in Figs.~\ref{Fig:V_Kt_alpha}(g)--\ref{Fig:V_Kt_alpha}(i), we plot the volumetric thermal expansivity $\alpha \equiv (1/V)(\partial V / \partial T)_P$ of the MS phase, along with its HS and LS counterparts ($\alpha_{HS}$ and $\alpha_{LS}$). During the SCO, $\alpha$ goes through an anomalous increase rather than just shift from $\alpha_{HS}$ to $\alpha_{LS}$. This can be understood via Eq.~(\ref{Eq:alpha}) below (see also Eq.~(S20) in SM \cite{SM})

\begin{equation}
\alpha V = n V_{LS}\alpha_{LS} + (1-n) V_{HS} \alpha_{HS} - (V_{HS}-V_{LS})\left(\frac{\partial n}{\partial T} \right)_P,
\label{Eq:alpha} 
\end{equation}

\bigskip
\noindent
which indicates that the anomaly of $\alpha$ mainly arises from $(\partial n/\partial T)_P$. By comparing $\alpha$ with $(\partial n/\partial T)_P$ shown in Figs.~\ref{Fig:n_dndP_dndT}(g)--\ref{Fig:n_dndP_dndT}(i), one can notice that the peaks of $\alpha$ and the dips of $(\partial n/\partial T)_P$ not only align with each other, but are also broadened and smeared by temperature in the same manner. Furthermore, our calculations also indicate that the anomalies of $K_T$ and $\alpha$ are quite significant even at low iron concentration. For $x=0.125$, $K_T$ drops by $47$\%, $31$\%, and $16$\% [Fig.~\ref{Fig:V_Kt_alpha}(d)], and $\alpha$ increases to $6.5$, $3.1$, and $2$ times larger [Fig.~\ref{Fig:V_Kt_alpha}(g)] in the SCO region at $T=300$, $600$, and $1200$ K, respectively. For $x=0.5$, $K_T$ drops by $77$\%, $61$\%, and $43$\% [Fig.~\ref{Fig:V_Kt_alpha}(e)], and $\alpha$ increases to $21$, $8.9$, and $4.6$ times larger [Fig.~\ref{Fig:V_Kt_alpha}(h)] at $T=300$, $600$, and $1200$ K, respectively. 

\bigskip

\begin{figure}[pt]
\begin{center}
\includegraphics[
]{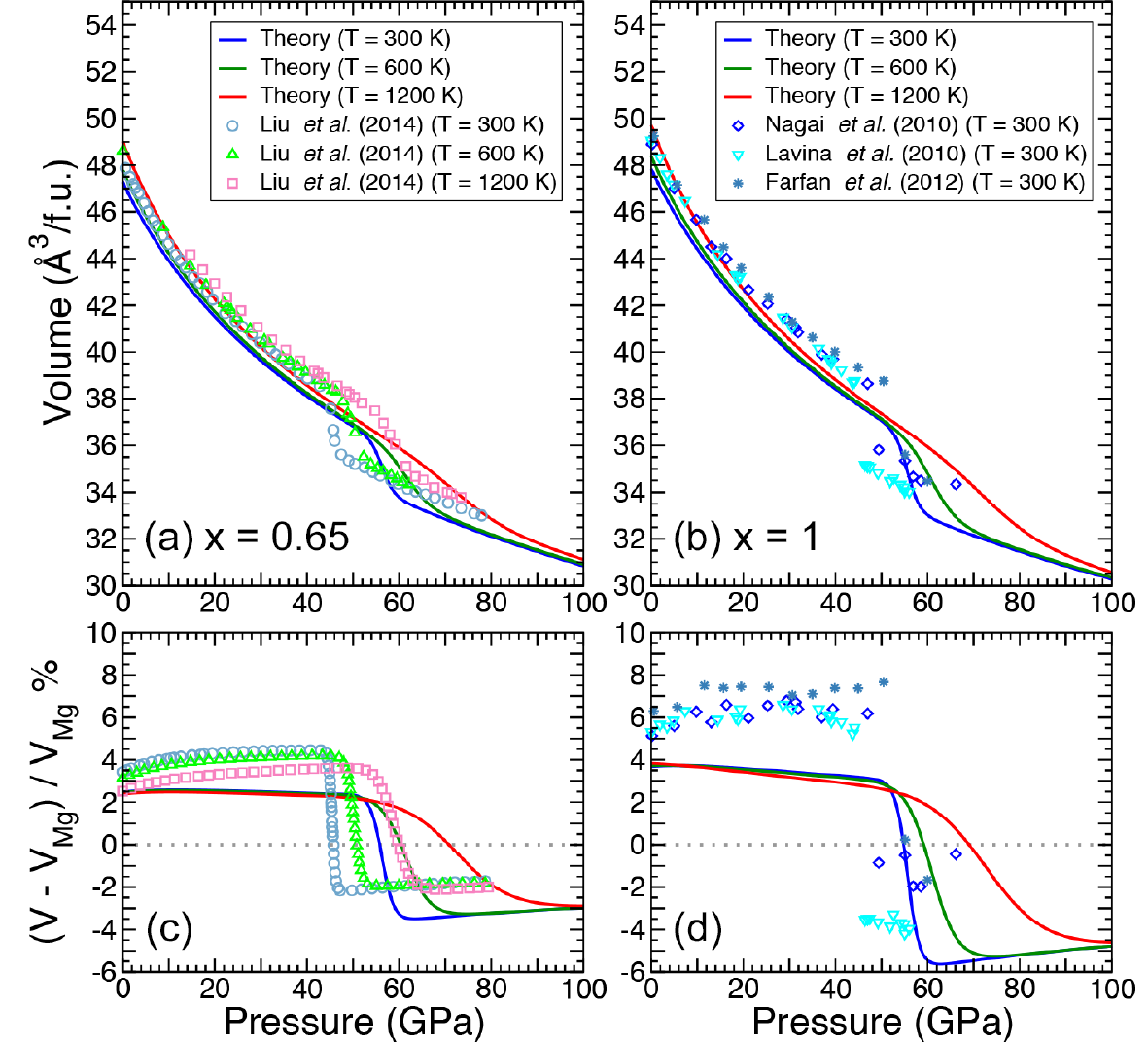}
\end{center}
\caption{(a,b) Compression curves of (Mg$_{1-x}$Fe$_x$)CO$_3$, and (c,d) relative volume differences between (Mg$_{1-x}$Fe$_x$)CO$_3$ and MgCO$_3$ ($V_{Mg}$) for $x=0.65$ and $1$. Solid lines denote our theoretical results; symbols denote experimental results \cite{Liu_2014_AmMin, Nagai10, Lavina_2010_PRB, Farfan12}.}
\label{Fig:exp_V_dV}
\end{figure}

Next, we compare our theoretical results with experiments for iron concentration $x=0.65$ by Liu \textit{et al}. \cite{Liu_2014_AmMin} and Fu \textit{et al}. \cite{Fu_2017_PRL}, and $x=1$ by Farfan \textit{et al}. \cite{Farfan12}, Lavina \textit{et al}. \cite{Lavina_2010_PRB}, and Nagai \textit{et al}. \cite{Nagai10}. The Gibbs free energy $G_i(P,T)$ of spin state $i$ ($i=$ HS or LS) for $x=0.65$ is obtained by interpolating the results of $x=0.5$ and $x=1$ (see Eq.~(S14) in SM \cite{SM}); from $G_i(P,T)$, the Gibbs free energy $G(P,T)$ and all thermal parameters of the MS phase can be determined. In Figs.~\ref{Fig:exp_V_dV}(a) and \ref{Fig:exp_V_dV}(b), compression curves $V(P)$ for $x=0.65$ and $x=1$ are shown. In our previous static calculation, theory underestimates the room-temperature equilibrium volume ($V_0$) by $\sim 4\%$ \cite{Hsu_2016_PRB}; in the present calculation with the inclusion of lattice vibration, such underestimate is reduced to $\sim 2$\%. For both iron concentrations, theoretical results are overall in good agreement with experiments. To better examine the volume anomaly, we plot the relative volume difference between (Mg$_{1-x}$Fe$_x$)CO$_3$ and MgCO$_3$ ($V_{Mg}$) for $x=0.65$ and $x=1$ in Figs.~\ref{Fig:exp_V_dV}(c) and \ref{Fig:exp_V_dV}(d), respectively. The computed and measured $V_{Mg}(P,T)$ \cite{Litasov_2008_PEPI} are adopted to plot the $(V-V_{Mg})/V_{Mg}$ curves for the theoretical and experimental results, respectively. For $x=1$ [Fig.~\ref{Fig:exp_V_dV}(d)], all three room-temperature experiments \cite{Nagai10, Lavina_2010_PRB, Farfan12} exhibit the same trend and show slight difference: (1) Overall, the SCO starts at as low as 45 GPa and finishes at as high as 60 GPa, (2) HS FeCO$_3$ is 5--8\% larger (in volume) than MgCO$_3$, and LS FeCO$_3$ is 2--4\% smaller than MgCO$_3$, and (3) a volume reduction of $\sim 9$\% occurs in the SCO region. In our calculation for $T=300$ K (indicated by the blue line), a volume reduction of $\sim 9$\% occurs in the SCO region 52--62 GPa, in good agreement with experiments. It should be pointed out that four different experiments are adopted for this comparison (three for FeCO$_3$; one for MgCO$_3$), and each experiment has its own systematic error. As can be observed, the measured FeCO$_3$ volumes in these experiments differ by $\sim 2$\%. Likewise, the measured MgCO$_3$ volume, which is used as the reference $V_{Mg}$ for experiments, may also have an uncertainty of $\sim 2$\%. Considering this factor, the apparent discrepancy between the theoretical and experimental results is in fact within the uncertainty of experiments. For $x=0.65$, [Fig.~\ref{Fig:exp_V_dV}(c)], our calculation is also in good agreement with the experiment by Liu \textit{et al}. \cite{Liu_2014_AmMin}. A volume reduction of $\sim 6.5$\% and the broadening of the SCO region with increasing temperature can be observed in both the theoretical and experimental results. On the other hand, the computed spin-transition pressures and SCO regions are $\sim 10$ GPa higher and 5--15 GPa wider, respectively, than the experimental results. The wider SCO region predicted by theory may be caused by a few factors, including the spatial distribution of Fe atoms and the modeling of the MS phase. As detailed in SM, we consider the MS phase as a solid solution of the HS and LS states \cite{SM}. Other modeling can lead to different spin-transition width, as shown in molecular-dynamics (MD) calculations for (Mg,Fe)O \cite{Holmstrom15}.

\bigskip

\begin{figure}[pt]
\begin{center}
\includegraphics[
]{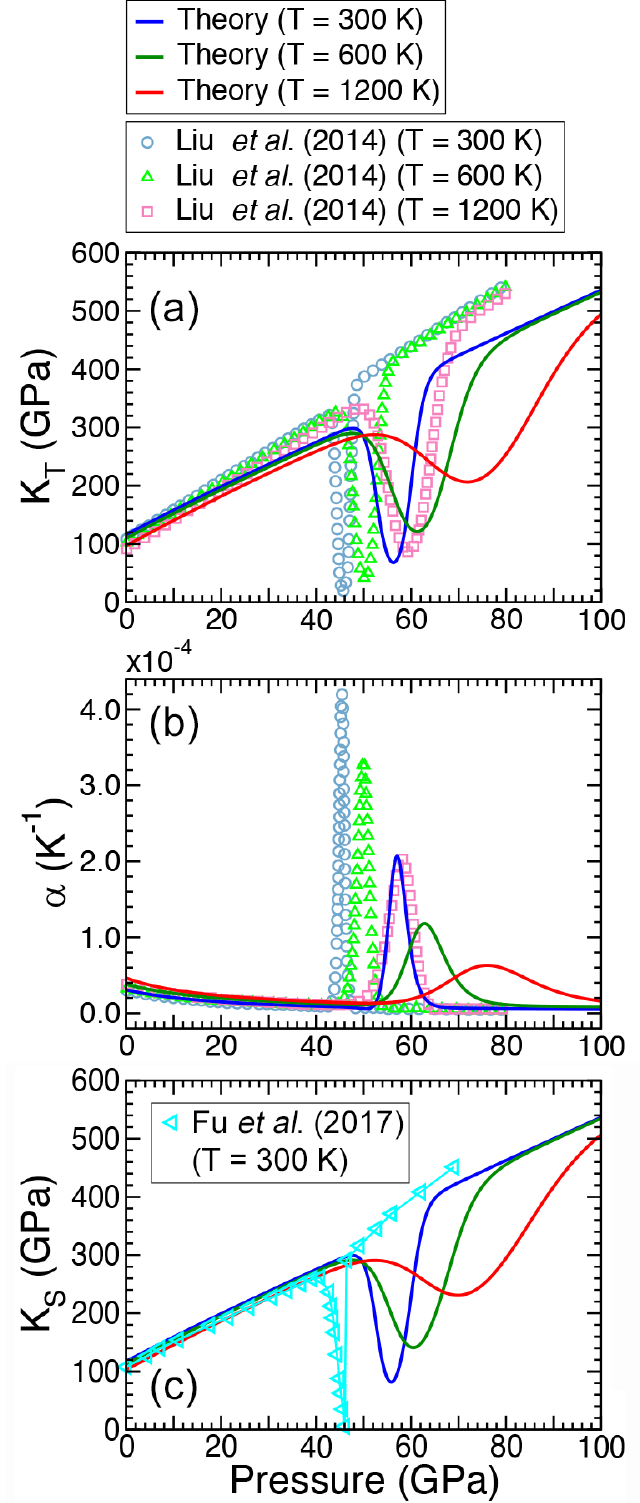}
\end{center}
\caption{(a) Isothermal bulk modulus $K_T$, (b) volumetric thermal expansivity $\alpha$, and (c) adiabatic bulk modulus $K_S$ of (Mg$_{0.35}$Fe$_{0.65}$)CO$_3$. Solid lines denote our theoretical results; symbols denote experimental results \cite{Liu_2014_AmMin, Fu_2017_PRL}. Noted that Liu \textit{et al}. obtained $K_T$ and $\alpha$ by the fitted EoS \cite{Liu_2014_AmMin}, not by direct measurement.}
\label{Fig:exp_Kt_alpha_Ks}
\end{figure}

\begin{figure}[pt]
\begin{center}
\includegraphics[
]{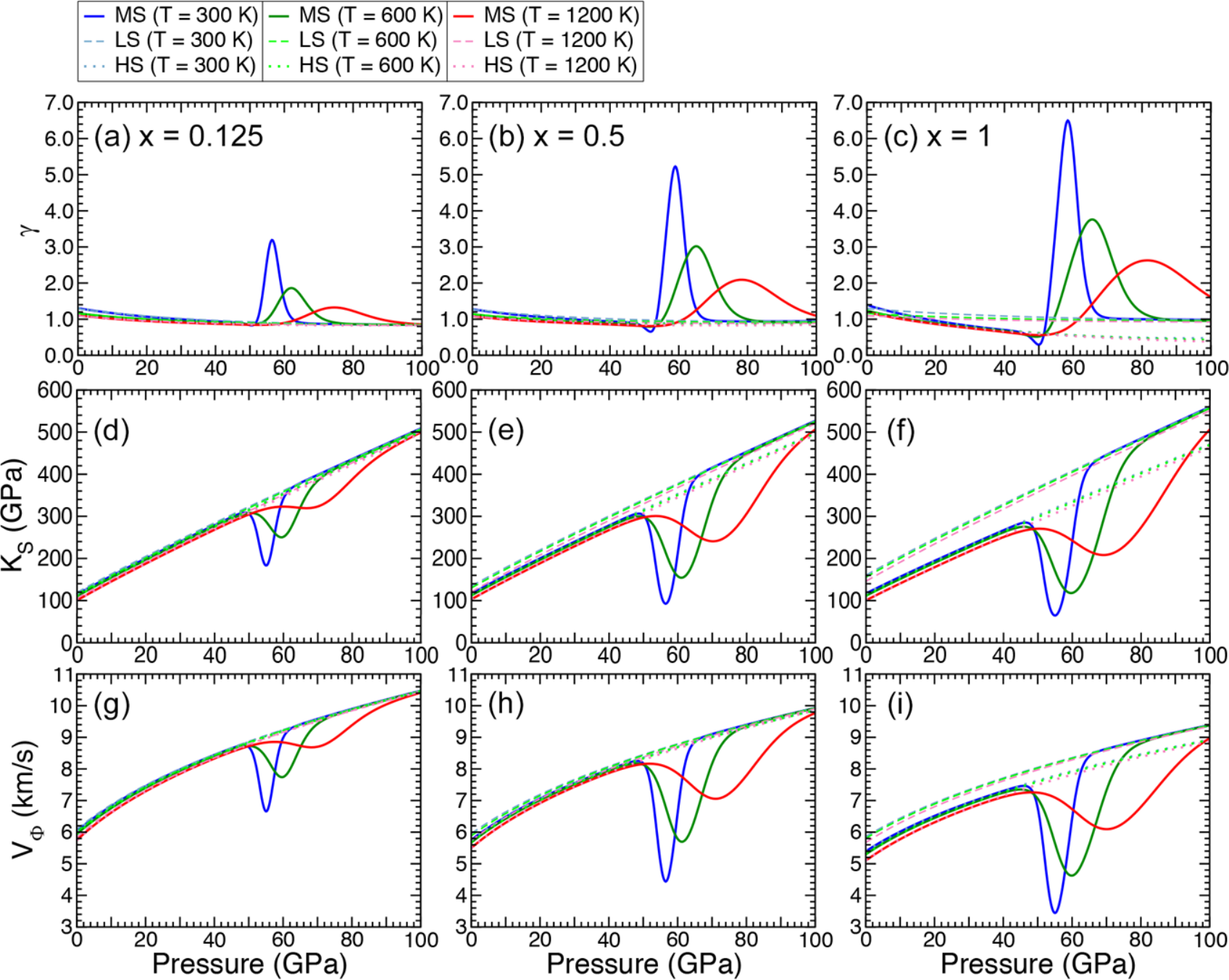}
\end{center}
\caption{(a--c) Thermodynamic Gr\"uneisen parameter $\gamma$, (d--f) adiabatic bulk modulus $K_S$, and (g--i) bulk sound velocity $V_{\Phi}$ of (Mg$_{1-x}$Fe$_x$)CO$_3$ for $x=0.125$, $0.5$, and $1$, respectively. Solid, dotted, and dashed lines denote our theoretical results for the MS, HS, and LS states, respectively.}
\label{Fig:gamma_Ks_Vphi}
\end{figure}

In Fig.~\ref{Fig:exp_Kt_alpha_Ks}, we compare the computed and measured bulk modulus $K_T$, volumetric thermal expansivity $\alpha$, and adiabatic bulk modulus $K_S \equiv -V(\partial P / \partial V)_S$ of (Mg$_{0.35}$Fe$_{0.65}$)CO$_3$ (see Fig.~\ref{Fig:gamma_Ks_Vphi} for the calculation of $K_S$). Overall, theoretical and experimental results are in agreement. As shown in Figs.~\ref{Fig:exp_Kt_alpha_Ks}(a) and \ref{Fig:exp_Kt_alpha_Ks}(b), anomalies of $K_T$ and $\alpha$ observed in the experiment by Liu \textit{et al}. \cite{Liu_2014_AmMin} are 25--50\% and $\sim 100$\% larger (in magnitude) than the theoretical results, respectively, despite that theory and experiment give the same volume anomalies of $\sim 6.5$\% (Fig.~\ref{Fig:exp_V_dV}). The main reason is that the SCO region observed by Liu \textit{et al}. is narrower than the theoretical results, namely, $(\partial n/\partial P)_T$ and $(\partial n/\partial T)_P$ observed by Liu \textit{et al}. have greater magnitudes, leading to greater anomalies in $K_T$ and $\alpha$, respectively [see Eqs.~(\ref{Eq:Kt}) and (\ref{Eq:alpha})]. As to $K_S$, the theoretical result for $T=300$ K is in excellent agreement with the room-temperature experiment by Fu \textit{et al}. \cite{Fu_2017_PRL} before the SCO ($P \lesssim 41$ GPa), while the anomaly observed in the experiment is slightly narrower and $\sim 15$\% larger than the theoretical result. Interestingly, in principle, $K_S$ should be larger than $K_T$ [see later in Eq.~(\ref{Eq:Ks})], but the measured $K_S$ [Fig.~\ref{Fig:exp_Kt_alpha_Ks}(c)] and $K_T$ [Fig.~\ref{Fig:exp_Kt_alpha_Ks}(a)] show otherwise. Such inconsistency between different experiments indicates that the uncertainties of experimental results may be larger than they seem.
 
\bigskip 


In Fig.~\ref{Fig:Cp_Cv}, we show our \textit{predictive} calculations for the constant-pressure ($C_P$) and constant-volume ($C_V$) heat capacities of (Mg$_{1-x}$Fe$_x$)CO$_3$ at high $P$--$T$ conditions. Their HS/LS counterparts ($C_P^{HS/LS}$ and $C_V^{HS/LS}$) are also plotted. The computed $C_P$ for FeCO$_3$ ($x=1$) at $T=300$ K [Fig.~\ref{Fig:Cp_Cv}(c)] is in good agreement with the room-temperature measurement \cite{Robie_1984_AmMin}. For all iron concentrations, the HS state has slightly larger heat capacities than the LS state ($C_P^{HS} > C_P^{LS}$; $C_V^{HS} > C_V^{LS}$), especially at lower temperature. This can be understood via the VDOS of ferromagnesite: HS Fe atoms vibrate with lower frequencies than LS Fe atoms [Figs.~\ref{Fig:VDOS_nPT}(b) and \ref{Fig:VDOS_nPT}(c)]. As the temperature increases to $1200$ K, such difference becomes negligible, even for FeCO$_3$ ($x=1$) [Figs.~\ref{Fig:Cp_Cv}(c) and \ref{Fig:Cp_Cv}(f)]. During the SCO, $C_P$ undergoes anomalous increases of $\sim 6$\%, $\sim 24$\%, and $\sim 45$\% for iron concentrations $x=0.125$, $0.5$, and $1$, respectively [Figs.~\ref{Fig:Cp_Cv}(a)--\ref{Fig:Cp_Cv}(c)]. Remarkably, the anomaly of $C_P$ retains its magnitude without smearing out at high temperature, in contrast to the anomalies of bulk modulus and thermal expansivity (Figs.~\ref{Fig:V_Kt_alpha} and \ref{Fig:exp_Kt_alpha_Ks}). This characteristic of $C_P$ can be understood via Eq.~(\ref{Eq:Cp}) below (see also Eq.~(S24) in SM \cite{SM}),

\begin{equation}
\begin{aligned}
C_P & \equiv T \left( \frac {\partial S}{\partial T} \right)_P \\
& = n C_P^{LS} + (1-n) C_P^{HS} + \
T(S_{LS}-S_{HS})\left( \frac {\partial n}{\partial T}\right)_P \
+ (G_{LS}-G_{HS}) \left( \frac {\partial n}{\partial T}\right)_{P}.
\label{Eq:Cp}
\end{aligned}
\end{equation}

\bigskip
\noindent
For $C_P$, the maximum of the anomaly occurs at around the spin phase boundary, namely, when $\Delta G_{LS} \equiv G_{LS}-G_{HS} \approx 0$. Therefore, when $C_P$ reaches its maximum,

\begin{equation}
C_P \approx n C_P^{LS} + (1-n) C_P^{HS} + \
T(S_{LS}-S_{HS})\left( \frac {\partial n}{\partial T}\right)_P, 
\label{Eq:CpMax}
\end{equation}

\bigskip
\noindent
which indicates that the maximum anomaly of $C_P$ is mainly determined by $T(\partial n/\partial T)_P$ rather than $(\partial n/\partial T)_P$. Since the smearing of $(\partial n/\partial T)_P$ with increasing temperature [Figs.~\ref{Fig:n_dndP_dndT}(g)--\ref{Fig:n_dndP_dndT}(i)] is now compensated by multiplying with $T$, the anomaly of $C_P$ remains prominent at high temperature. As to $C_V$ [Figs.~\ref{Fig:Cp_Cv}(d)--\ref{Fig:Cp_Cv}(f)], the anomalous increases in the SCO region are significantly smaller than those of $C_P$; outside of the SCO region, $C_V$ and $C_P$ are nearly the same. This can be understood via Eq.~(\ref{Eq:Cv}) below

\begin{equation}
C_V \equiv T \left(\frac {\partial S}{\partial T}\right)_V \\
= C_P-T V \alpha^2 K_T,
\label{Eq:Cv} 
\end{equation}

\bigskip
\noindent
where the term $T V \alpha^2 K_T$, also plotted in Figs.~\ref{Fig:Cp_Cv}(d)--\ref{Fig:Cp_Cv}(f), is small outside of the SCO region and exhibits an anomalous increase in the SCO region. 

\bigskip


A couple of implications can be drawn from our analysis for the heat capacity $C_P$. First, among the currently available experiments, the SCO region reported in Ref.~\onlinecite{Liu_2014_AmMin} by Liu \textit{et al}. is among the narrowest, providing possible upper limits for the magnitudes of $(\partial n/\partial P)_T$ and $(\partial n/\partial T)_P$. Based on the comparison of thermal expansivity $\alpha$ in Fig.~\ref{Fig:exp_Kt_alpha_Ks}(b), $(\partial n/\partial T)_P$ observed in Ref.~\onlinecite{Liu_2014_AmMin} can be twice as large as our theoretical result. Since the anomaly of $C_P$ is determined by $T(\partial n/\partial T)_P$ [Eq.~(\ref{Eq:CpMax})], we estimate that the anomaly of $C_P$ in (Mg$_{1-x}$Fe$_x$)CO$_3$ during SCO would be 6--12\%, 24--48\%, and 45--90\% for iron concentration $x=0.125$, $0.5$, and $1$, respectively. Such a significant change of $C_P$ during SCO may affect the temperature of subducting slabs. Second, as mentioned in Sec. I, either thermal diffusivity $D \equiv \kappa / \rho C_P$ or thermal effusivity $e \equiv \sqrt{\kappa \rho C_P}$ are measured ($\rho$ is density) in TR-based experiments. To accurately extract thermal conductivity ($\kappa$) from TR-based experiments, accurate $C_P$ is a necessary input. In practice, since $C_P$ at high $P$--$T$ conditions are not easily available, estimated $C_P$ (often a constant) are adopted without considering the anomaly of $C_P$ during SCO \cite{Chao_2019_JGR, Hsieh18, Hsieh17}. For example, in Ref.~\onlinecite{Chao_2019_JGR}, thermal conductivity of (Mg$_{0.22}$Fe$_{0.78}$)CO$_{3}$ is extracted from thermal effusivity measured via TDTR. By assuming a constant $C_{P}$, the authors report an anomalous increase of $\kappa$ during SCO, from $11$ to $45$ W/m/K (increasing by $310$\%). Based on our discussion of Fig.~\ref{Fig:Cp_Cv}, however, $C_{P}$ has an anomalous increase of $35$--$70$\% during SCO. By taking the anomaly of $C_P$ into account, the anomalous increase of $\kappa$ should be smaller, namely, from $11$ to $26$--$33$ W/m/K. Our results thus call for further examinations of thermal conductivities extracted from TR-based experiments for Fe-bearing minerals, including ferromagnesite \cite{Chao_2019_JGR}, ferropericlase \cite{Ohta17, Hsieh18}, and bridgmanite \cite{Hsieh17, Okuda19}, given the significant anomaly of $C_P$ accompanying SCO.

\bigskip


With $C_P$ and $C_V$ obtained, a few more thermal parameters can be determined, including thermodynamic Gr\"uneisen parameter $\gamma$, adiabatic bulk modulus $K_S$, and bulk sound velocity $V_{\Phi}$. In general, the thermodynamic Gr\"uneisen parameter $\gamma \equiv V \alpha K_T / C_V$ of a material marginally changes with pressure and temperature, as can be observed from pure HS and LS (Mg$_{1-x}$Fe$_{x}$)CO$_3$ [Figs.~\ref{Fig:gamma_Ks_Vphi}(a)--\ref{Fig:gamma_Ks_Vphi}(c)]. In the SCO region, however, $\gamma$ exhibits an anomalous increase, which smears out as the temperature increases, similar to thermal expansivity $\alpha$. Noticeably, even at low iron concentration $x=0.125$, anomalies of $\gamma$ are still prominent: $\sim 260$\%, $\sim 116$\%, and $\sim 60$\% at $T=300$, $600$, and $1200$ respectively [Fig.~\ref{Fig:gamma_Ks_Vphi}(a)]. For adiabatic bulk modulus $K_S$ [Figs.~\ref{Fig:gamma_Ks_Vphi}(d)--\ref{Fig:gamma_Ks_Vphi}(f)], its anomalous softening is similar to that of the isothermal bulk modulus $K_T$ [Figs.~\ref{Fig:V_Kt_alpha}(d)--\ref{Fig:V_Kt_alpha}(f)], given that 

\begin{equation}
K_S \equiv -V \left(\frac{\partial P}{\partial V}\right)_S \\
= K_T \frac{C_P}{C_V} = K_T (1+ \gamma \alpha T). 
\label{Eq:Ks} 
\end{equation}

\bigskip
\noindent
Outside of the SCO region, $C_P \approx C_V$ (Fig.~\ref{Fig:Cp_Cv}), therefore, $K_S \approx K_T$; in the SCO region, $1 < C_P/C_V \lesssim 1.5$ (Fig.~\ref{Fig:Cp_Cv}), so the dips of $K_S$ are slightly shallower than those of $K_T$. Since the bulk sound velocity $V_{\Phi} \equiv \sqrt{K_S/\rho}$, the anomaly of $V_{\Phi}$ resembles that of $K_S$ [Figs.~\ref{Fig:gamma_Ks_Vphi}(g)--\ref{Fig:gamma_Ks_Vphi}(i)]. Based on the phonon gas model, thermal conductivity $\kappa = \frac{1}{3} C_P V_{\Phi} l = \frac{1}{3} C_P V_{\Phi}^2 \tau$, where $l$ and $\tau$ are the phonon mean free path and phonon scattering time, respectively. Anomalies of $C_P$ and $V_{\Phi}$ in the SCO region thus directly contribute to the anomalous change of thermal conductivity (see Ref.~\onlinecite{Wu18} for a discussion on ferropericlase). Calculations for $\kappa$ and $\tau$ from the first principles, however, are beyond the scope of this paper.

\section{Conclusion}

In this work, we perform first-principles LDA+$U_{sc}$ calculations to study the iron spin crossover and thermal properties of ferromagnesite (Mg$_{1-x}$Fe$_{x}$)CO$_3$ up to high pressure ($P=100$ GPa) and temperature ($T=1200$ K). Our calculations show that throughout a wide range of iron concentration ($0 < x \leq 1$) , the spin phase diagram of ferromagnesite remains nearly the same. The spin transition pressure $P_t$, the width of the SCO region, and their increase with temperature are barely affected by iron concentration. Our calculations also show that the thermal properties of (Mg$_{1-x}$Fe$_{x}$)CO$_3$ are drastically altered by SCO, including anomalous reduction of volume, anomalous softening of bulk modulus, and anomalous increase of thermal expansivity. These results are overall in good agreement with experiments. Our calculations also predict anomalous increases of heat capacity and thermodynamic Gr\"uneisen parameter during SCO. Remarkably, the anomaly of constant-pressure heat capacity $C_P$ remains prominent at high temperature without smearing out, in contrast to the anomalies of bulk modulus, thermal expansivity, and bulk sound velocity. This result suggests significant change of thermal conductivity during SCO; it also calls for further examinations of the results obtained from TR-based experiments, as inaccurate $C_P$ has been adopted to extract thermal conductivity. Our results further suggest that SCO may significantly affect the thermal properties and temperature of subducting slabs, given that several minerals abundant in subducting slabs undergo SCO in the lower-mantle pressure range, including ferromagnesite, ferropericlase, the NAL, and the CF phases.

\bigskip

\textbf{Acknowledgments} This work is supported by the Ministry of Science and Technology of Taiwan under Grants No.~MOST~107-2112-M-008-022-MY3 and 107-2119-M-009-009-MY3. Z.W. is supported by National Natural Science Foundation of China (41925017). Calculations were performed primarily at National Center for High-performance Computing (NCHC) of Taiwan and partly at the Supercomputing Center of University of Science and Technology of China.

\newpage

\newpage

\newpage

\newpage

\newpage

\newpage

\newpage

\newpage

\end{document}



\title{Supplemental Material for ``Anomalous thermal properties and spin crossover of ferromagnesite (Mg,Fe)CO$_3$"}

\author{Han Hsu}
\affiliation{Department of Physics, National Central University, Taoyuan City 32001, Taiwan}
\author{C. Crisostomo}
\affiliation{Department of Physics, National Central University, Taoyuan City 32001, Taiwan}
\author{Wenzhong Wang}
\affiliation{School of Earth and Space Sciences, University of Science and Technology of China, Hefei, Anhui 230026, P.~R. China}
\author{Zhongqing Wu}
\affiliation{School of Earth and Space Sciences, University of Science and Technology of China, Hefei, Anhui 230026, P.~R. China}

\date{\today}

\maketitle

\section{Thermodynamic model}

In this paper, we study lattice vibration and thermal properties of ferromagnesite (Mg$_{1-x}$Fe$_x$)CO$_3$ by computing phonon spectra ($\omega^i_{\nu\bold q}$) of spin state $i$ ($i=$ HS, IS, or LS) at various volumes ($V$) for various iron concentrations ($x$). Within the quasi-harmonic approximation (QHA), the vibrational free energy $F^{vib}_i$ of spin state $i$ at volume $V$ is written as

\begin{equation}
F^{vib}_i(T,V)=\sum_{\nu\bold q} \hbar \omega^i_{\nu\bold q}/2 + k_{B}T \sum_{\nu\bold q} \ln [ 1- \exp (-\hbar\omega^i_{\nu\bold q}/k_{B}T)].
\label{Eq:Fvib} 
\end{equation}

\bigskip
\noindent
With the inclusion of static contribution, we write $F^{stat+vib}_i(T,V) = F^{vib}_i(T,V) + E^{stat}_i(V)$, where $E^{stat}_i(V)$ is the static LDA+$U_{sc}$ energy of spin state $i$. Given that $P = -\partial F^{stat+vib}_i/ \partial V$, the compression curve $V_i(P,T)$ of spin state $i$ can be computed. The Gibbs free energy $G_i(P,T)$ of spin state $i$ is thus written as

\begin{equation}
G_i(P,T)=G^{vib+stat}_i-k_{B}Tx\ln[2S^{el}_{i}+1],
\label{Eq:Gi}
\end{equation}

\bigskip
\noindent
where $G^{vib+stat}_i=F^{vib+stat}_i(T,V)+PV_i(P,T)$ includes the vibrational and static contribution, $S^{el}_{i}$ is the total electron spin of Fe in spin state $i$, and the term $k_{B}x \ln[2S^{el}_{i} + 1]$ is the magnetic entropy. For ferromagnesite, $S^{el}_{HS} = 2$, and $S^{el}_{LS} = 0$. 

\bigskip

At $T \neq 0$, ferromagnesite goes through a mixed-spin (MS) phase/state, in which all spin states coexist. The fraction of spin state $i$ in the MS phase is written as $n_{i}(P,T)$. Here we consider the MS phase as a solid solution of all spin states, and the Gibbs free energy of the MS phase is written as

\begin{equation}
G(P,T)=\sum\limits_{i} n_i(P,T)G_i(P,T) - TS^{mix},
\label{Eq:G-1}
\end{equation}

\bigskip
\noindent
where $S^{mix}$ is the mixing entropy of the solid solution, given by

\begin{equation}
S^{mix} = -k_Bx \sum\limits_{i} n_i \ln n_i.
\label{Eq:Smix-1}
\end{equation}

\bigskip
\noindent
As described in the main text, the IS state of ferromagnesite is highly \textit{unfavorable}, and the IS fraction $n_{IS}$ is negligible. Effectively, $n_{IS}=0$, and $n_{LS}+n_{HS}=1$. For convenience, we write $n \equiv n_{LS}$ and $n_{HS}=1-n$. Using this notation, the Gibbs free energy and mixing entropy in Eqs.~(\ref{Eq:G-1}) and (\ref{Eq:Smix-1}) are now rewritten as

\begin{eqnarray}
G(n,P,T) &=& nG_{LS} + (1-n)G_{HS} - TS^{mix},
\label{Eq:G} \\
S^{mix} &=& - k_Bx [n \ln n + (1-n) \ln (1-n)].
\label{Eq:Smix}
\end{eqnarray}

\bigskip
\noindent
The total derivative of $G(n,P,T)$ is

\begin{equation}
dG= \left( \frac{\partial G}{\partial n} \right)_{P,T} dn + \left( \frac{\partial G}{\partial P} \right)_{n,T} dP + \left( \frac{\partial G}{\partial T} \right)_{n,P} dT,
\label{Eq:dG}
\end{equation}

\bigskip
\noindent
with

\begin{eqnarray}
\left( \frac{\partial G}{\partial n} \right)_{P,T} &=& \
G_{LS} - G_{HS} - T \left( \frac {\partial S^{mix}}{\partial n}\right)_{P,T}, 
\label{Eq:dGdn} \\
\left( \frac {\partial S^{mix}}{\partial n}\right)_{P,T} &=& \
k_Bx \ln \left( \frac {1-n}{n} \right),
\label{Eq:dSmixdn} \\
\left( \frac{\partial G}{\partial P} \right)_{n,T} \
&=& n \left( \frac{\partial G_{LS}}{\partial P} \right)_T \
+ (1-n) \left( \frac{\partial G_{HS}}{\partial P} \right)_T,
\label{Eq:dGdP} \\
\left( \frac{\partial G}{\partial T} \right)_{n,P} \
&=& n \left( \frac{\partial G_{LS}}{\partial T} \right)_P \
+ (1-n) \left( \frac{\partial G_{HS}}{\partial T} \right)_P - S^{mix}. 
\label{Eq:dGdT}
\end{eqnarray}

\bigskip
\noindent
At equilibrium, the Gibbs free energy at a given $(P,T)$ is minimized. To ensure vanishing $dG$ in Eq.~(\ref{Eq:dG}) with $dP=dT=0$, we must require $(\partial G / \partial n )_{P,T} = 0$. This requirement, along with Eqs.~(\ref{Eq:dGdn}) and (\ref{Eq:dSmixdn}), gives the LS fraction

\begin{equation}
n=\frac{1}{1+\exp(\Delta G_{LS}/k_{B}Tx)},
\label{Eq:n}
\end{equation}

\bigskip
\noindent
where $\Delta G_{LS} \equiv G_{LS}-G_{HS}$. Equivalently, we can also write

\begin{equation}
n=\frac{1}{1+(2S^{el}_{HS}+1)\exp(\Delta G^{vib+stat}_{LS}/k_{B}Tx)},
\label{Eq:n}
\end{equation}

\bigskip
\noindent
where $\Delta G^{vib+stat}_{LS} \equiv G^{vib+stat}_{LS}-G^{vib+stat}_{HS}$.

\bigskip
In this work, the Gibbs free energy $G_i(P,T)$ of spin state $i$ for iron concentrations $x=0.125$, $0.5$, and $1$ are computed via LDA+$U_{sc}$ and QHA. For other iron concentration, the Gibbs free energy $G_i(P,T)$ are computed via interpolation. For example, for $x=0.65$, we write 

\begin{equation}
G_i^{(x=0.65)}(P,T) = 0.7 \times G_i^{(x=0.5)}(P,T) + 0.3 \times G_i^{(x=1)}(P,T), 
\end{equation}

\bigskip
\noindent
from which the LS fraction $n(P,T)$ and thermal parameters of (Mg$_{0.35}$Fe$_{0.65}$)CO$_{3}$ can be derived. 

\section{Thermal parameters}

In Eq.~(\ref{Eq:G}), the Gibbs free energy $G$ of the MS phase is expressed in terms of $G_{LS}$, $G_{HS}$, and the LS fraction $n$. Similarly, thermal parameters of the MS phase can also be expressed in terms of their HS/LS counterparts and $n$. Given that $(\partial G / \partial n)_{P,T}=0$ at equilibrium,

\begin{eqnarray}
\left( \frac{\partial G}{\partial P} \right)_T &=& \left( \frac{\partial G}{\partial P} \right)_{n,T} + \left( \frac{\partial G}{\partial n} \right)_{P,T} \left( \frac{\partial n}{\partial P} \right)_T = \left( \frac{\partial G}{\partial P} \right)_{n,T},\\
\left( \frac{\partial G}{\partial T} \right)_P &=& \left( \frac{\partial G}{\partial T} \right)_{n,P} + \left( \frac{\partial G}{\partial n} \right)_{P,T} \left( \frac{\partial n}{\partial T} \right)_P = \left( \frac{\partial G}{\partial T} \right)_{n,P}.
\end{eqnarray}

\bigskip
\noindent
Combining the above results with Eqs.~(\ref{Eq:dGdP}) and (\ref{Eq:dGdT}), expressions for the volume and entropy ($S$) of the MS phase can be obtained 

\begin{eqnarray}
V &=& \left( \frac{\partial G}{\partial P} \right)_T \
= \left( \frac{\partial G}{\partial P} \right)_{n,T}\
= n V_{LS} + (1-n) V_{HS},
\label{Eq:V} \\
S &=& -\left( \frac{\partial G}{\partial T} \right)_P \
= -\left( \frac{\partial G}{\partial T} \right)_{n,P}\
= n S_{LS} + (1-n) S_{HS} + S^{mix}.
\label{Eq:S}
\end{eqnarray}

\bigskip
\noindent
Expressions for isothermal bulk modulus $K_T \equiv  -V(\partial P/ \partial V)_T$ and volumetric thermal expansivity $\alpha \equiv -(1/V)(\partial V/ \partial T)_P$ can be obtained by taking the derivatives of volume [Eq.~(\ref{Eq:V})] with respect to pressure and temperature, respectively  

\begin{eqnarray}
\frac{V}{K_T} &=& -\left( \frac{\partial V}{\partial P} \right)_T \ 
= n \frac{V_{LS}}{K_{T}^{LS}} + (1-n) \frac{V_{HS}}{K_{T}^{HS}} + (V_{HS}-V_{LS})\left(\frac{\partial n}{\partial P} \right)_T,
\label{Eq:Kt} \\
\alpha V &=& \left( \frac{\partial V}{\partial T} \right)_P \
= n V_{LS}\alpha_{LS} + (1-n) V_{HS} \alpha_{HS} - (V_{HS}-V_{LS})\left(\frac{\partial n}{\partial T} \right)_P.
\label{Eq:alpha} 
\end{eqnarray}

\bigskip

To find out the expression for the constant-pressure heat capacity $C_P \equiv T(\partial S/ \partial T)_P$ of the MS phase, we first derive $(\partial S/ \partial T)_P$ via Eq.~(\ref{Eq:S})

\begin{equation}
\left( \frac {\partial S}{\partial T} \right)_P = \\
n \left( \frac {\partial S_{LS}}{\partial T} \right)_P + \\
(1-n) \left( \frac {\partial S_{HS}}{\partial T} \right)_P + \\
(S_{LS} - S_{HS}) \left( \frac {\partial n}{\partial T} \right)_P + \\
\left( \frac {\partial S^{mix}}{\partial T}\right)_P.
\label{Eq:dSdT} 
\end{equation}

\bigskip
\noindent
The term $(\partial S^{mix}/\partial T)_P$ in Eq.~(\ref{Eq:dSdT}) can be expressed in terms of the LS fraction $n$ via Eq.~(\ref{Eq:dSmixdn})

\begin{equation}
\left( \frac {\partial S^{mix}}{\partial T}\right)_P = \
\left( \frac {\partial S^{mix}}{\partial n}\right)_{T,P}\
\left( \frac {\partial n}{\partial T}\right)_{P} = \
k_Bx \left(\frac{\partial n}{\partial T}\right)_P\ln\left(\frac{1-n}{n} \right).
\label{Eq:dSmixdT}
\end{equation} 

\bigskip
\noindent
Again, since $(\partial G / \partial n)_{P,T}=0$ at equilibrium, Eqs.~(\ref{Eq:dSmixdT}), (\ref{Eq:dGdn}), and (\ref{Eq:dSmixdn}) lead to

\begin{equation}
\left( \frac {\partial S^{mix}}{\partial T}\right)_P = \\
\frac{G_{LS}-G_{HS}}{T} \left( \frac {\partial n}{\partial T}\right)_{P}
\label{Eq:dSmixdT-2}
\end{equation}

\bigskip
\noindent
With Eqs.~(\ref{Eq:dSdT})--(\ref{Eq:dSmixdT-2}), $C_P$ can be expressed as below

\begin{equation}
\begin{aligned}
C_P & \equiv T \left( \frac {\partial S}{\partial T} \right)_P \\
&= n C_P^{LS} + (1-n) C_P^{HS} + \
T(S_{LS}-S_{HS})\left( \frac {\partial n}{\partial T}\right)_P \
+ k_BTx \left( \frac {\partial n}{\partial T}\right)_P \ln \left(\frac {1-n}{n}\right) \\
& = n C_P^{LS} + (1-n) C_P^{HS} + \
T(S_{LS}-S_{HS})\left( \frac {\partial n}{\partial T}\right)_P \
+ (G_{LS}-G_{HS}) \left( \frac {\partial n}{\partial T}\right)_{P}.
\label{Eq:Cp}
\end{aligned}
\end{equation}

\bigskip
During spin crossover, thermal parameters of the MS phase exhibit anomalous changes. The maximum of the anomalous change occurs at around the HS/LS phase boundary, namely, when $G_{LS}-G_{HS} \approx 0$, or equivalently, $n \approx 0.5$ (and thus $\ln[(1-n)/n] \approx 0$). Therefore, when the anomaly of $C_P$ reaches the maximum,

\begin{equation}
C_P \approx n C_P^{LS} + (1-n) C_P^{HS} + \
T(S_{LS}-S_{HS})\left( \frac {\partial n}{\partial T}\right)_P, 
\end{equation}

\bigskip
\noindent
which indicates that the maximum anomaly of $C_P$ is mainly determined by $T(\partial n/\partial T)_P$ rather than $(\partial n/\partial T)_P$. Despite that $(\partial n/\partial T)_P$ smears out as the temperature increases, $T(\partial n/\partial T)_P$ remains prominent at high $T$. Consequently, the anomaly of $C_P$ remains prominent at high temperature without smearing out, in contrast to the anomalies of thermal expansivity $\alpha$ and isothermal bulk modulus $K_T$.

\section{Supplemental results}

In Fig.~4 of the main text, compression curve of pure HS and LS (Mg$_{1-x}$Fe$_x$)CO$_3$ ($x=0.125$, $0.5$, and $1$) are plotted for $T=300$, $600$, and $1200$ K. As described in Sec. II, these curves are obtained by fitting our calculation results with the 3rd BM EoS. Zero-pressure volume ($V_0$), bulk modulus ($K_0$), and $K_0' \equiv (\partial K_T / \partial P)|_{P=0}$ of these curves are are tabulated in Tables SI--SIII. 

\bigskip

As mentioned in the main text, for $P \gtrsim 50$, (Mg$_{1-x}$Fe$_x$)CO$_3$ with $x \geq 0.65$ is unstable at $T \gtrsim 1300$ K. For lower Fe concentration ($x$), (Mg$_{1-x}$Fe$_x$)CO$_3$ may remain stable to higher temperature at high pressure. In Fig.~S1, compression curves and thermal parameters of (Mg$_{0.5}$Fe$_{0.5}$)CO$_3$ up to $T=2000$ K are shown. Notably, the anomalous increase of $C_P$ remains prominent at $T=2000$ K [Fig.~S1(d)]. 

\newpage
\begin{table*}
\caption{3rd BM EoS parameters of HS and LS (Mg$_{1-x}$Fe$_x$)CO$_3$ with $x=0.125$.}
\begin{ruledtabular}
\begin{tabular}{ccccc}
 ($x=0.125$) & & $T=300$ K & $T=600$ K & $T=1200$ K\\
 \hline
\multirow{3}{*}{HS} & $V_0$ (\AA$^3$/f.u.) & 46.3877 & 46.8896 & 48.1348
 \\
 & $K_0$ (GPa) & 113.4475 & 107.3533 & 94.5500 \\
 & $K_0'$ & 4.5603 & 4.6225 & 4.7510 \\
\hline
\multirow{3}{*}{LS} & $V_0$ (\AA$^3$/f.u.) & 45.5082 & 45.9834 & 47.1579 \\
 & $K_0$ (GPa) & 117.9147 & 111.9551 & 99.5224 \\
 & $K_0'$ & 4.5728 & 4.6247 & 4.7255 \\
\end{tabular}
\end{ruledtabular}
\end{table*}

\newpage
\begin{table*}
\caption{3rd BM EoS parameters of HS and LS (Mg$_{1-x}$Fe$_x$)CO$_3$ with $x=0.5$.}
\begin{ruledtabular}
\begin{tabular}{ccccc}
 ($x=0.5$) & & $T=300$ K & $T=600$ K & $T=1200$ K\\
 \hline
\multirow{3}{*}{HS} & $V_0$ (\AA$^3$/f.u.) & 47.0388 & 47.5244 & 48.7139
 \\
 & $K_0$ (GPa) & 115.4896 & 109.7049 & 97.5242 \\
 & $K_0'$ & 4.4718 & 4.5297 & 4.6519 \\
\hline
\multirow{3}{*}{LS} & $V_0$ (\AA$^3$/f.u.) & 43.7371 & 44.1470 & 45.1579 \\
 & $K_0$ (GPa) & 133.2505 & 127.2482 & 114.6184 \\
 & $K_0'$ & 4.5658 &  4.6155 & 4.7135 \\
\end{tabular}
\end{ruledtabular}
\end{table*}

\newpage
\begin{table*}
\caption{3rd BM EoS parameters of HS and LS (Mg$_{1-x}$Fe$_x$)CO$_3$ with $x=1.0$.}
\begin{ruledtabular}
\begin{tabular}{ccccc}
 ($x=1.0$) & & $T=300$ K & $T=600$ K & $T=1200$ K\\
 \hline
\multirow{3}{*}{HS} & $V_0$ (\AA$^3$/f.u.) & 47.8354 & 48.3708 & 49.7102
 \\
 & $K_0$ (GPa) & 115.3718 & 108.3411 & 93.7692 \\
 & $K_0'$ & 4.1648 & 4.2413 & 4.3805 \\
\hline
\multirow{3}{*}{LS} & $V_0$ (\AA$^3$/f.u.) & 41.1905 & 41.5350 & 42.3864 \\
 & $K_0$ (GPa) & 158.8778 & 152.6019 & 139.3280 \\
 & $K_0'$ & 4.5974 &  4.6358 & 4.7095 \\
\end{tabular}
\end{ruledtabular}
\end{table*}

\newpage
\begin{figure*}[pt]
\begin{center}
\includegraphics[
]{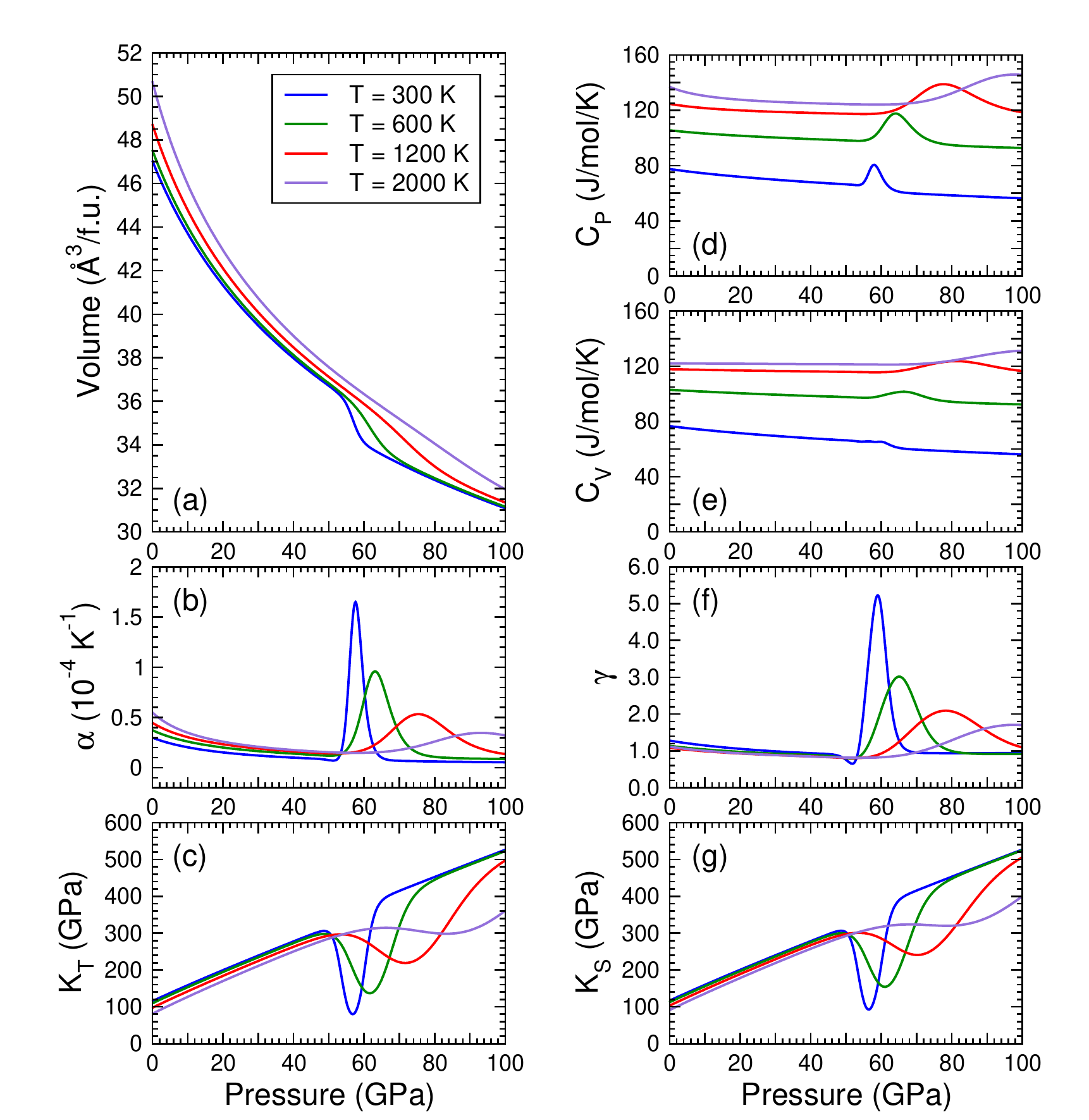}
\end{center}
\caption{Compression curves and thermal parameters of (Mg$_{0.5}$Fe$_{0.5}$)CO$_3$ (in the MS state) at various temperatures. See main text for the definitions of thermal parameters shown in panels (b)--(g).}
\label{Fig:S1}
\end{figure*}